\documentclass[12pt]{iopart}

\usepackage{graphicx,bbm}

\def\tr{\,{\rm tr}\,}
\def\ket#1{|#1\rangle}

\def\ave#1{\langle #1 \rangle}
\def\ii{{\rm i}}

\def\tit#1{{\em #1},}
\def\etal#1{#1}

\begin{document}

\title{Dephasing-induced diffusive transport in anisotropic Heisenberg model}

\author{Marko \v Znidari\v c}
\address{Department of Physics, Faculty of Mathematics and Physics,
  University of Ljubljana, Ljubljana, Slovenia}

\date{\today}

\begin{abstract}
We study transport properties of anisotropic Heisenberg model in a disordered magnetic field and in the presence of dephasing due to external degrees of freedom. Without dephasing the model can display, depending on parameter values, the whole range of possible transport regimes: ideal ballistic conduction, diffusive, or ideal insulating behavior. We show that the presence of dephasing induces normal diffusive transport in a wide range of parameters. We also analyze the dependence of spin conductivity on the dephasing strength. In addition, by analyzing the decay of spin-spin correlation function we find a long-range order for finite chain sizes. All our results for a one-dimensional spin chain at infinite temperature can be equivalently rephrased for strongly-interacting disordered spinless fermions.
\end{abstract}



\section{Introduction}

Theory of free fermions has been very successful in explaining properties of many condensed matter systems. Sometimes though, in particular in low dimensional systems, the picture of isolated noninteracting fermions is an oversimplifications and one has to include additional interactions. Integrable model of free fermions can be upgraded by taking into account several effects: (i) interaction -- fermions can interact with each other, (ii) disorder -- fermions can experience different local eigenenergies and, (iii) coupling --  fermions can interact with external degrees of freedom. A non-perturbative inclusion of any of these effects greatly complicates the analysis forcing one to use various successful phenomenological theories~\cite{Gantmakher}. There is nevertheless a desire to understand properties of such strongly-interacting many-body disordered systems from the first principles, that is starting from the governing equations of motion. 

\begin{figure}[!h]
\centerline{\includegraphics[angle=0,scale=0.4]{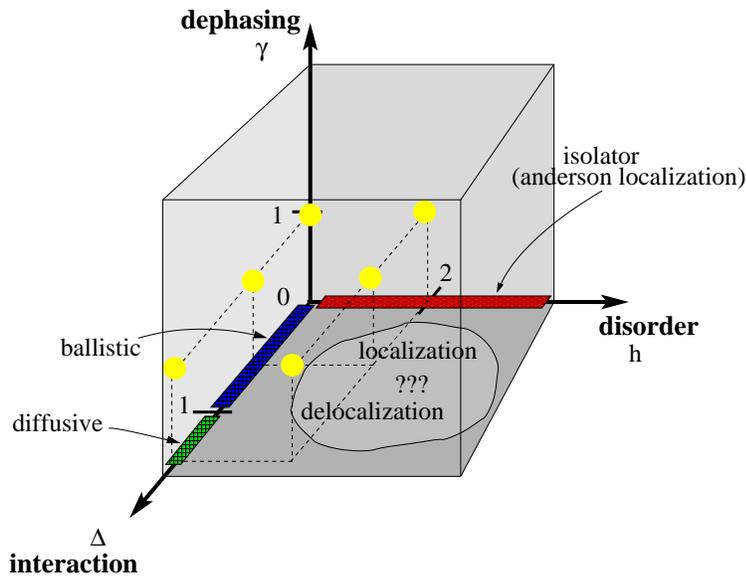}}
\caption{(Color online) Parameter space of the Heisenberg model in the presence of interaction $\Delta$, disorder with strength $h$ and dephasing $\gamma$. Rigorous transport properties are known only on axes $\Delta=0$ or $h=0$, both with $\gamma=0$. In the present work we show that dephasing ($\gamma \neq 0$, yellow balls) induces diffusive transport.}
\label{fig:diagram}
\end{figure}
In the present work we shall study transport properties of a one-dimensional (1D) antiferromagnetic Heisenberg model at infinite temperature in the presence of all the above mentioned effects: interaction, disorder and coupling to the environment. To our knowledge this is a first study of a large many-body system in such setting by using microscopic equations of motion. In terms of Pauli matrices the 1D antiferromagnetic spin-1/2 Heisenberg model reads
\begin{equation}
H=\sum_{j=1}^{n-1} (\sigma_j^{\rm x} \sigma_{j+1}^{\rm x} +\sigma_j^{\rm y} \sigma_{j+1}^{\rm y}+ \Delta \sigma_j^{\rm z} \sigma_{j+1}^{\rm z})+ \sum_{j=1}^n h_j \sigma_j^{\rm z}.
\label{eq:H}  
\end{equation}
Magnetic field strength $h_j$ at site $j$ is chosen according to a uniform distribution in the interval $[-h,h]$. Even though a clean Heisenberg model without disordered magnetic field is explicitly solvable by the Bethe ansatz and has been studied for several decades, its spin transport properties are far from being understood, for an overview see~\cite{Brenig:07}. By using the Jordan-Wigner transformation it can be mapped to a system of strongly interacting spinless fermions, with $\Delta$ being the interaction strength. Therefore, the problem of spin transport in the Heisenberg model is equivalent to that of particle (charge) transport in a system of interacting fermions. In the absence of coupling to the environment, rigorous statements are available only for special cases without anisotropy (interaction), $\Delta=0$, or in the case of no disorder, $h=0$, see Fig.~\ref{fig:diagram}. For $\Delta=0$ the model is equivalent to a system of non-interacting fermions and can be described in terms of single-particle states that exhibit exponential localization in the presence of nonzero disorder, $h\neq 0$, resulting in a perfect insulating behavior due to Anderson localization~\cite{Anderson}. If disorder is zero the Heisenberg model displays ballistic spin transport in a gapless phase for $\Delta<1$~\cite{Zotos:97}, while it is likely a diffusive spin conductor for $\Delta>1$~\cite{Meisner:03,Samir:04,JSTAT:09,Langer:09,Steinigeweg:09}. This last result about diffusive transport in an integrable model, based mainly on numerical investigations, is somewhat unexpected and still lacks a deeper understanding, see though~\cite{Buragohain,Affleck:09} for theoretical arguments supporting diffusive transport. As soon as interaction $\Delta$ and disorder $h$ are both nonzero things get more complicated because one has to deal with a genuine many-body physics, for single-particle treatment see, e.g.,~\cite{singleparticle,Gurvitz:00}. There are in fact opposing claims in the literature on whether the localization is destroyed or preserved in the presence of strong interaction; some predict a phase transition from an insulating to a conducting regime at high temperature~\cite{Basko:06}, other numerical simulations, albeit on small systems, suggest localization even at an infinite temperature~\cite{Oganesyan:07,PRB:08} or even ideal conducting behavior~\cite{Karahalios:09}.   

Other non-equilibrium properties, like dynamics in various quantum quenches, has also been studied intensely in recent years, see~\cite{Barmettler} and references therein. For a recent study in 3D see~\cite{weimer:08}. Besides its theoretical importance, Heisenberg model is also experimentally realized in spin-chain materials~\cite{experiment}. It is argued that an anomalously high spin conductivity of the Heisenberg model is required to explain large measured heat conductivity. While a detailed engineering of interactions in condensed-matter systems is probably out of question, experiments with cold gases in optical lattices could in near future lead to controlled experiments with strongly correlated many-body systems~\cite{cold}. It is therefore important to get a better understanding of the influence of the interplay between interaction, disorder and external coupling on the dynamics in such systems.

\section{The model}

We are going to study transport by numerically solving dynamical equations of motion, finding a non-equilibrium stationary state (NESS) to which the system converges after long time. Once we calculate NESS we can evaluate various expectation values in that stationary non-equilibrium state. The dynamics of the system will be described by the Lindblad master equation~\cite{lindblad},
\begin{equation}
\frac{{\rm d}}{{\rm d}t}{\rho}=\ii [ \rho,H ]+ {\cal L}^{\rm bath}(\rho)+{\cal L}^{\rm deph}(\rho).
\label{eq:Lin}
\end{equation}
Bath superoperator ${\cal L}^{\rm bath}$ takes into account the coupling to unequal heat baths at both ends, inducing a non-equilibrium situation, while ${\cal L}^{\rm deph}$ represents the effect of dephasing caused by the coupling to some external degrees of freedom. Even though description in terms of Lindblad equation, originating in quantum optics, has not been employed very often in condensed matter physics, it has been recently used in several works studying spin transport in the Heisenberg model~\cite{Wichterich:07,Michel:08,JSTAT:09,Langer:09,Steinigeweg:09a}. While a number of assumptions is involved in the derivation of the Lindblad master equation from unitary evolution of a combined system~\cite{Breuer}, Markovian approximation can be justified because we are interested in the stationary state reached after long time, e.g., much longer than the bath correlation time. Apart from coupling to the bath the only non-unitary effect we take into account is dephasing. At high temperatures, which is the regime studied in the present work, dephasing is the dominant environmental effect. We are going to solve master equation using time-dependent density matrix renormalization group (tDMRG) method enabling us to study by an order of magnitude larger chains~\cite{JSTAT:09,Langer:09} than is possible with other methods. Dephasing present in our simulations in fact acts stabilizing on the numerical method enabling us to simulate relaxation in chains of up-to $512$ spins. 
\begin{figure}[!t]
\centerline{\includegraphics[angle=0,scale=0.28]{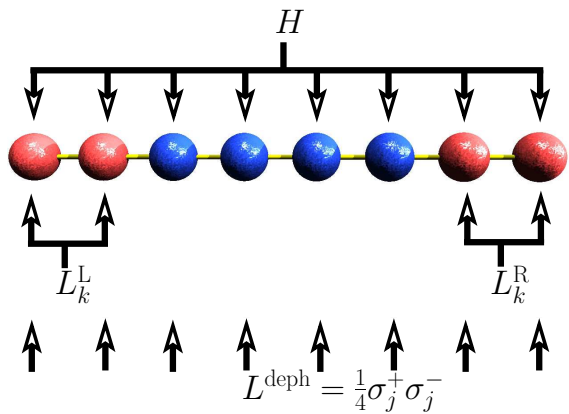}}
\caption{(Color online) Hamiltonian part acts on all spins, Lindblad part for the baths acts on the boundary two spins, while the dephasing part acts independently on each spin.}
\label{fig:mpsL}
\end{figure}

Let us describe various terms in the Lindblad equation (\ref{eq:Lin}). The Lindblad bath superoperator ${\cal L}^{\rm bath}$ affects only two border spins at the left and right end of the chain and involves four Lindblad operators $L_{k=1,2,3,4}^{\rm L}$ at the left end and four $L_{k=1,2,3,4}^{\rm R}$ at the right end,
\begin{equation}
{\cal L}^{\rm bath}(\rho)=\sum_k \left( [ L_k^{\rm L,R} \rho,L_k^{{\rm L,R}\dagger} ]+[ L_k^{\rm L,R},\rho L_k^{{\rm L,R}\dagger} ] \right).
\end{equation}
Lindblad operators $L_k^{\rm L,R}$ are chosen in such way that if they would be the only term present in the master equation, the NESS, i.e., the state $\rho(t \to \infty)$, would be equal to the grand canonical state on two border spins. They are obtained by targeting the grand canonical state at an infinite temperature, $\rho_{\rm grand.} \sim \exp{(-\mu_{\rm L,R} \Sigma^{\rm z})}$, $\Sigma^{\rm z}=\sum_k \sigma_k^{\rm z}$, on two boundary spins, see~\cite{JSTAT:09,ergod} for more details. By choosing different potential $\mu_{\rm L}$ and $\mu_{\rm R}$ at the left and the right end, a NESS with a nonzero spin current is obtained. The bath operators are such that for $\mu_{\rm L}=\mu_{\rm R}$ they induce the correct equilibrium grand canonical state in the bulk of an interacting chaotic spin chain~\cite{ergod}. Dephasing part ${\cal L}^{\rm deph}$, due to, for instance, coupling with phonons, involves only one Lindblad operator~\footnote{The same ${\cal L}^{\rm deph}$ is obtained for $L=\frac{1}{2}\sigma_j^{\rm z}$.} $L=\frac{1}{4}\sigma_j^+ \sigma_j^-$ for each spin site $j$,
\begin{equation}
{\cal L}^{\rm deph}(\rho)=\frac{\gamma}{16} \sum_j \left( [ \sigma_j^+ \sigma_j^- \rho, \sigma_j^+ \sigma_j^- ]+[ \sigma_j^+ \sigma_j^-,\rho\, \sigma_j^+ \sigma_j^- ] \right),
\end{equation}
where the sum over $j$ runs over spin sites, $\sigma^\pm=\sigma^{\rm x} \pm {\rm i}\, \sigma^{\rm y}$, and $\gamma$ is a dephasing strength. Dephasing (also called phase damping or phase-flip in quantum information) causes an exponential decay of the off-diagonal elements in the diagonal basis of $\sigma^{\rm z}$ (in fermionic language this corresponds to the off-diagonal decay in the number basis). It might be interesting to note~\cite{private} that the dynamics of operators in the Heisenberg model in the presence of dephasing, i.e., evolution by the Lindblad equation (\ref{eq:Lin}), can be exactly mapped to the evolution of pure states in a one-dimensional non-hermitian Hubbard model with complex on-site repulsion $({\rm i}U) n_{j \uparrow} n_{j \downarrow}$.

Recently, the influence of dephasing on transport has been studied in short disordered networks modelling the light-harvesting complex~\cite{Mohseni:08,Plenio:08,Rebentrost:09,Caruso:09}, where the presence of dephasing can enhance network's ability to transmit excitations. 

\section{Non-equilibrium steady states}
\begin{figure}[!th]
\centerline{\includegraphics[scale=0.48,angle=-90]{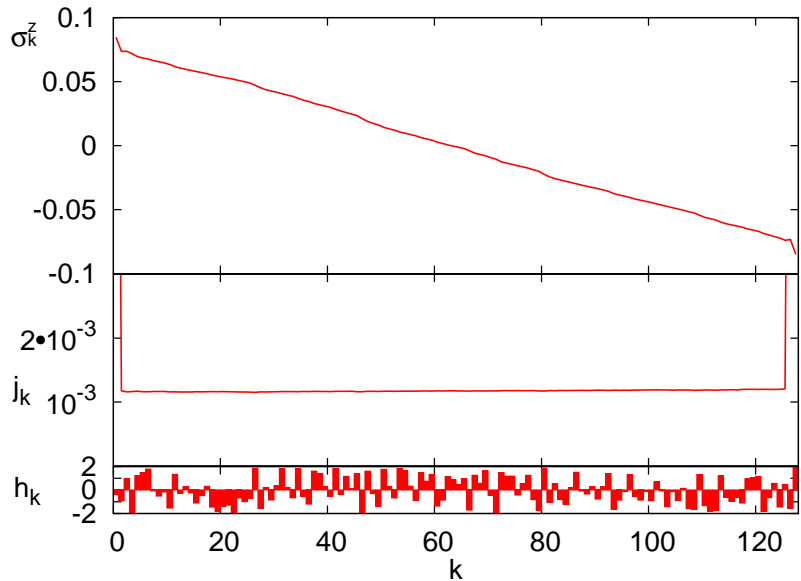}}
\caption{Spin and current profiles in the stationary state of a chain of $n=128$ spins with $\Delta=1.5$, disorder strength $h=2$ and dephasing $\gamma=1$. Bottom panel shows the size of magnetic field at individual sites.}
\label{fig:example}
\end{figure}
First we are going to show expectation values of some simple observables in NESS obtained as an infinite-time solution of the Lindblad equation (\ref{eq:Lin}). We choose bath potentials on left and right ends as $\mu_{\rm L,R}=\pm 0.1$ (same throughout the paper) and numerically solve the Lindblad equation, starting with some arbitrary initial condition. Details of our numerical implementation of tDMRG can be found in the Appendix. After sufficiently long time the state $\rho(t)$ converges to a time-independent NESS. This ``relaxation'' time after which we converge to the NESS depends on the bath details, i.e., it is a combined property of the central system and of the coupling. It increases with the increasing chain size $n$. For each run we carefully check that the simulation time has been long enough so that the convergence to the NESS is indeed reached. We show in Fig.~\ref{fig:example} the expectation value of local magnetization $\sigma_k^{\rm z}$ and of local spin current $j_k=2(\sigma_k^{\rm x} \sigma_{k+1}^{\rm y}-\sigma_k^{\rm y} \sigma_{k+1}^{\rm x})$ in a NESS of a disordered Heisenberg model with $n=128$ spins. One obtains a linear spin (magnetization) profile, as it is expected for a diffusive conductor, as well as a homogeneous spin current $j_k$ throughout the chain. Our numerical experience tells us that the homogeneity of the spin current can be used as a rather sensitive indicator of the convergence to NESS as well as of truncation errors. 

\subsection{Spin conductivity}
\begin{figure}[!h]
\centerline{\includegraphics[scale=0.5,angle=-90]{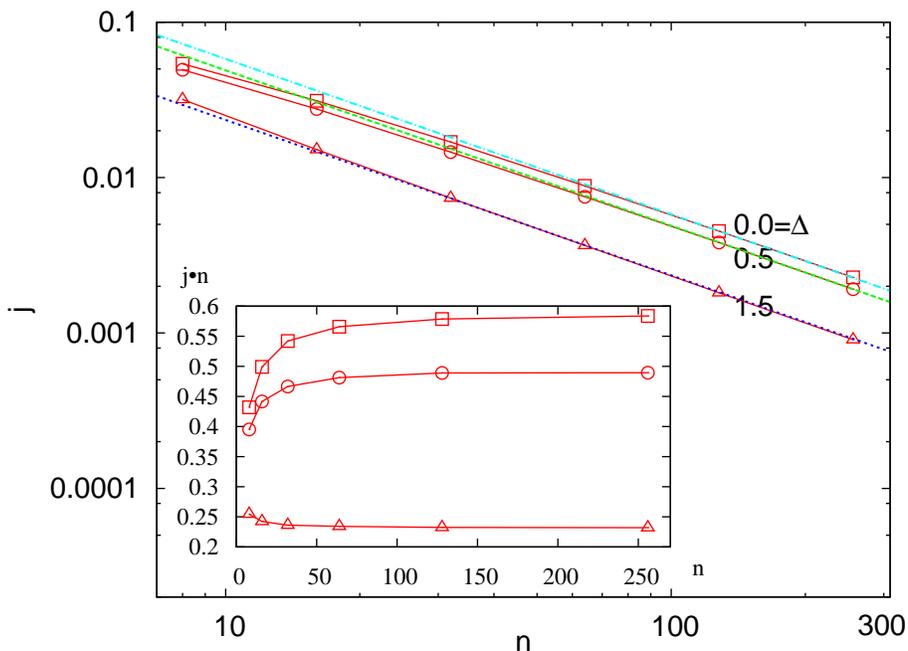}}
\caption{Scaling of the spin current with system's size in a clean system, $h=0$ and $\gamma=1$, at a fixed driving potential. For all anisotropies the presence of dephasing induces normal (diffusive) transport, signaled by the scaling $j \sim 1/n$ (straight lines in the main plot). In the inset the convergence of the product $j\cdot n$ to its asymptotic value is shown.}
\label{fig:scalingh0}
\end{figure}

To determine whether the system is a normal or an anomalous conductor one has to study the scaling of the spin current with the system size. First, we are going to study clean system without any disorder, $h=0$. Dephasing strength is always $\gamma=1$ and three different anisotropies $\Delta=0,0.5$, and $\Delta=1.5$ are used. Keeping chemical potentials fixed we have calculated the spin current in the NESS for systems between $n=8$ and $n=256$ spins long. Results are shown in Fig.~\ref{fig:scalingh0}. We can see that at $\gamma=1$, regardless of the value of the anisotropy, the current always scales as $\sim 1/n$ with the system size, meaning that the system is diffusive (i.e., normal) conductor. The asymptotic value of the product $j\cdot n$ determines the coefficient of spin conductivity $\kappa$ through $j=-\kappa (\ave{\sigma^{\rm z}_1}-\ave{\sigma^{\rm z}_n})/n$. For large $n$ we have $\ave{\sigma^{\rm z}_1}-\ave{\sigma^{\rm z}_n} \approx 0.153$ giving $\kappa=3.79, 3.20, 1.53$ for $\gamma=1$ and $\Delta=0,0.5,1.5$, respectively. With the increasing interaction $\Delta$ the conductivity decreases. For $\Delta=1.5$ and without dephasing and disorder, $\gamma=0$ and $h=0$, the coefficient of spin conductivity has been calculated in Refs.~\cite{JSTAT:09,Steinigeweg:09} to be $\kappa=2.30$ and is therefore larger than with dephasing. 

\begin{figure}[!h]
\centerline{\includegraphics[scale=0.5,angle=-90]{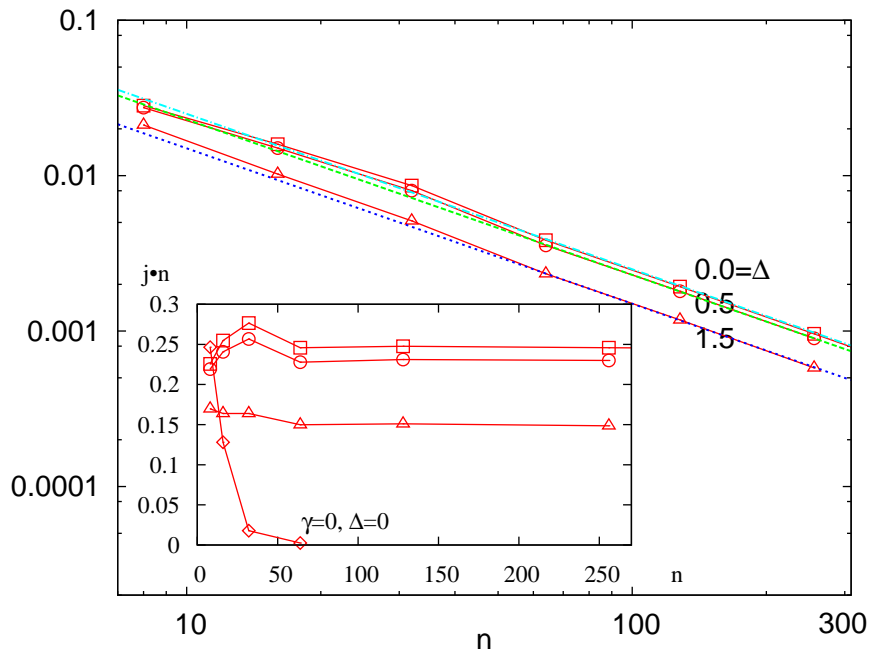}}
\caption{Same as Fig.~\ref{fig:scalingh0} for a disordered system, $h=2$ and $\gamma=1$. In the inset data for Anderson localization in the absence of dephasing and interaction, $\gamma=0,\, \Delta=0$, is also shown.}
\label{fig:scalingh2}
\end{figure}

Similar results are obtained also in the presence of disorder. In Fig.~\ref{fig:scalingh2} we show the scaling of the spin current with $n$ when each site, in addition to the dephasing with $\gamma=1$, experiences also a random magnetic field $h_j$ with strength $h=2$~\footnote{Because fluctuations between different realizations of disorder were found to be small for large $n$ all data shown are for a single realization of disorder.}. The difference in magnetizations between left and right ends is again almost independent of $\Delta$ and equal to $\ave{\sigma^{\rm z}_1}-\ave{\sigma^{\rm z}_n} \approx 0.153$, resulting in $\kappa=1.63, 1.50, 1.00$ for $\Delta=0,0.5,1.5$, respectively. Compared to the clean case disorder decreases conductivities, however, transport is still diffusive. Dephasing therefore breaks any possible many-body localization present at $\gamma=0$. In the inset of Fig.~\ref{fig:scalingh2} we also show data for the case of single-particle Anderson localization ($\gamma=0,\, \Delta=0$) for which the current exponentially decreases with $n$. 

Dephasing therefore induces normal transport irrespective of the anisotropy and disorder strength. Decreasing dephasing strength to $0$ one of course has to recover known behavior for $\gamma=0$ which is, depending on $\Delta$ and $h$ (see also Fig.~\ref{fig:diagram}), superconducting (for $\Delta < 1$ and $h=0$), diffusive (for $\Delta > 1$ and $h=0$) or insulating (for $\Delta=0$ and $h \neq 0$). To elucidate this transition we have studied how the coefficient of spin conductivity $\kappa$ changes as the dephasing strength $\gamma$ is varied. For each value of parameters $\Delta$ and $h$ we calculated NESS for sizes $n=64,128$ and $n=256$, from which we determined $\kappa$. Spin current in all cases scaled as $\sim 1/n$, the transition to asymptotics though happening at larger $n$ for smaller values of $\gamma$. 
\begin{figure}[!h]
\centerline{\includegraphics[scale=0.4,angle=-90]{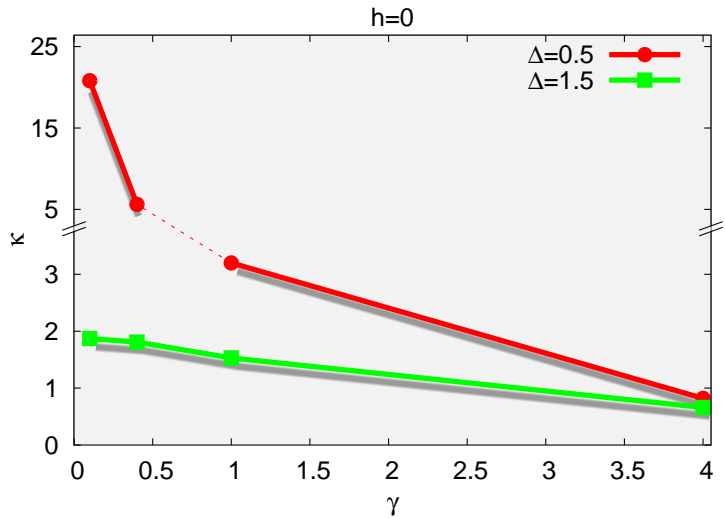}}
\caption{Dependence of the spin conductivity $\kappa$ on the dephasing strength $\gamma$ in the absence of magnetic field.}
\label{fig:kappah0}
\end{figure}
Dependence of $\kappa$ in the absence of disorder is shown in Fig.~\ref{fig:kappah0}. One can see that for $\Delta=0.5$ spin conductivity increases with decreasing $\gamma$. Divergence is scaling as $\kappa \sim 1/\gamma$ which is consistent with the superconducting (i.e., ballistic) limit at $\gamma=0$. For $\Delta=1.5$ though, $\kappa$ converges to a finite value as $\gamma \to 0$, in accordance with the normal transport in the Heisenberg model at $\Delta>1$ without dephasing. 

\begin{figure}[!h]
\centerline{\includegraphics[scale=0.4,angle=-90]{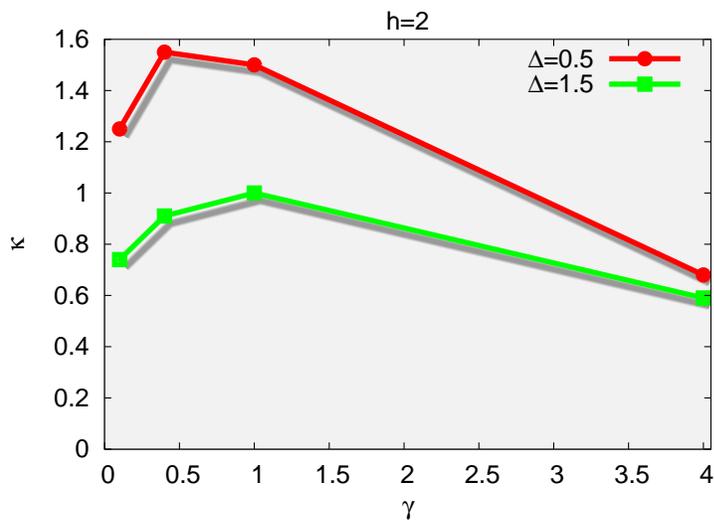}}
\caption{Dependence of the spin conductivity $\kappa$ on the dephasing strength $\gamma$ in the presence of disordered magnetic field with strength $h=2$.}
\label{fig:kappah2}
\end{figure}

In the presence of disorder the limit $\gamma \to 0$ corresponds to Anderson localization if $\Delta=0$ while for nonzero $\Delta$ things are not so clear. From our results seen in Fig.~\ref{fig:kappah2} we see that for small $\gamma$ spin conductivity decreases, however, the limit value $\kappa(\gamma=0)$ is rather difficult to predict. Spin conductivity reaches a maximum at some intermediate value of $\gamma$ and again decreases for large $\gamma$. Note that there is no qualitative difference between $\Delta=0.5$ and $\Delta=1.5$. A nontrivial maximum of conductivity is found also when studying dephasing in photosynthetic complex~\cite{Mohseni:08,Plenio:08,Rebentrost:09,Caruso:09}. 

\subsection{Correlation function}
\begin{figure}[!h]
\centerline{\includegraphics[scale=0.95]{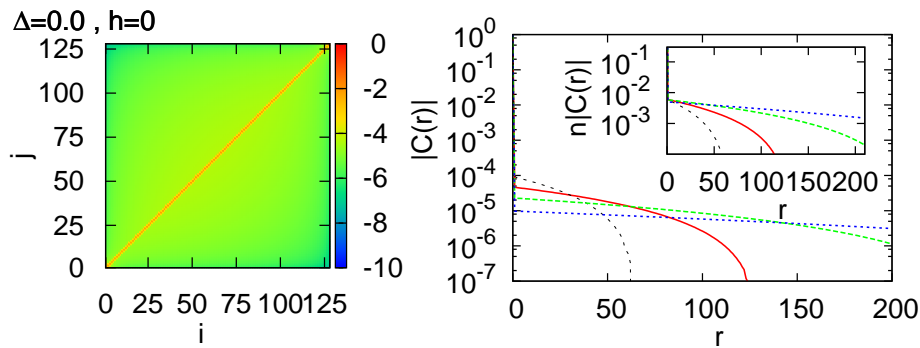}}
\caption{(Color online) Absolute value of the spin-spin correlation function $C(i,j)$ for NESS of a chain with $n=128$ spins and $h=0$, $\gamma=1$. Color code is for $\log_{10}{|C(i,j)|}$ (left panel). Right panel: cross-section along the diagonal, $C(r)=C(n/2-r/2,n/2+r/2)$; data are for $n=64,128,256$ and $n=512$ (left to right).}
\label{fig:corelh0}
\end{figure}
\begin{figure}[!h]
\centerline{\includegraphics[scale=0.95]{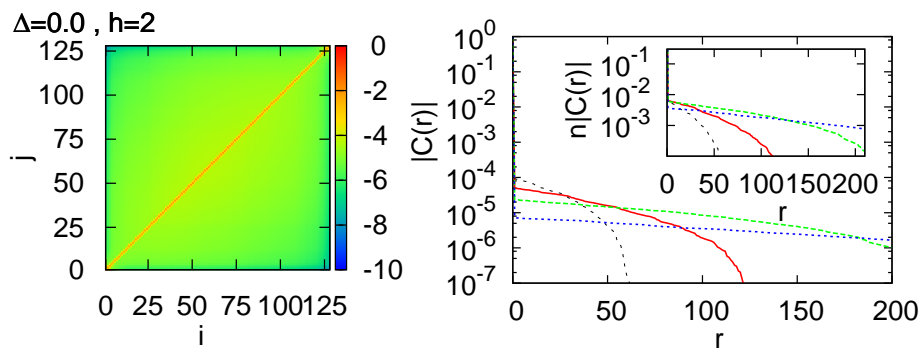}}
\caption{(Color online) The same as in Fig.~\ref{fig:corelh0}, but for a disordered model with $h=2$.}
\label{fig:corelh2}
\end{figure}

We have also calculated the spin-spin correlation function for non-equilibrium steady states from Figs.~\ref{fig:scalingh0} and~\ref{fig:scalingh2}. Results are similar for all values of $\Delta$ and we show only the case with $\Delta=0.0$, for which tDMRG is the most efficient. Spin correlation function $C(i,j)=\ave{\sigma_i^{\rm z}\sigma_j^{\rm z}}-\ave{\sigma_i^{\rm z}}\ave{\sigma_j^{\rm z}}$, shown in Figs.\ref{fig:corelh0} and~\ref{fig:corelh2}, in all cases has a plateau. This indicates the presence of a long-range order. Scaling analysis shows that the plateau in $C(i,j)$ scales with the system size as $\sim 1/n$ and with the potential difference as $\sim (\Delta \mu)^2=(\mu_{\rm L}-\mu_{\rm R})^2$. The same scaling is obtained also for $\Delta \neq 0$ (data not shown), however, because of less favorable scaling of the entanglement with $n$ we could reliably calculate $C(i,j)$ only for sizes $n \sim 128$. Note that decreasing values of the correlation function for increasing size make a precise calculation of $C(i,j)$ rather demanding. Scaling of the correlation function plateau $C \sim (\mu_{\rm L}-\mu_{\rm R})^2/n \sim j\cdot (\mu_{\rm L}-\mu_{\rm R})$ means that the long-range order is a purely non-equilibrium phenomenon. In equilibrium, when the spin current is zero, $j=0$, it goes to zero. In addition, correlations go to zero also in the thermodynamic limit $n \to \infty$, as one would expect for a true normal (diffusive) conductor. Similar scaling of the correlation plateau has been recently observed in a non-equilibrium phase transition in the XY spin chain in a homogeneous magnetic field, studied by exactly solving the corresponding master equation~\cite{arxiv:08,bojan:09}.

\section{Conclusion}
We have studied spin transport in the spin-1/2 Heisenberg model at infinite temperature in the presence of disorder and dephasing. By solving master equation for non-equilibrium driving potential we numerically obtained a non-equilibrium steady state and calculated expectation values of different observables. Studying the scaling of the spin current with system's size for systems of up-to $256$ spins we found that the dephasing always induces diffusive transport regardless of the anisotropy or disorder strength. The size of spin conductivity coefficient has a non-trivial maximum at intermediate values of the dephasing strength for a disordered model while it monotonically decreases with dephasing in the absence of disorder. Studying the spin-spin correlation function in a non-equilibrium steady state we have found the evidence for a long-range order at all anisotropies. The value of the correlation plateau scales as $\sim (\mu_{\rm L}-\mu_{\rm R})^2/n$, meaning that the correlations disappear in the thermodynamic limit as well as in the equilibrium situation. It would be interesting to study more in detail how the transition from diffusive transport to either superconducting, diffusive or insulating behavior happens as the dephasing strength is decreased to zero. More studies would be also needed to clarify the situation with a non-zero anisotropy and without dephasing, which corresponds to strongly interacting fermions.

\ack

I would like to thank Toma\v z Prosen for discussions and acknowledge grants J1-2208 and P1-0044 of the Slovenian Research Agency.

\appendix
\section*{Appendix}
\label{sec:appendix}
\setcounter{section}{1}

Here we briefly describe our implementation of tDMRG, for more details see~\cite{vidal}.

\subsection{Matrix product formulation}
Any state $\rho$ from a Hilbert space ${\cal H}$ which is a tensor product of $n$ local spaces of dimension $d$, ${\cal H}={\cal H}_{\rm loc.}^{\otimes n}$, ${\rm dim}({\cal H}_{\rm loc.})=d$, can be written in a matrix product operator (MPO) form, where expansion coefficients are expressed in terms of product of matrices. If we denote basis states of ${\cal H}_{\rm loc.}$ at $j$-th site by $\sigma_j^\nu, \nu=0,\ldots,d-1$, we can write
\begin{equation}
\ket{\rho}=\sum_{\nu_i} x^\dagger M_1^{\nu_1} M_2^{\nu_2}\cdots M_n^{\nu_n} x \ket{\sigma_1^{\nu_1} \sigma_2^{\nu_2} \cdots \sigma_n^{\nu_n}}.
\label{eq:MPO}
\end{equation}
For each site $j$ we have $d$ matrices $M_j^{\nu_j}$, each of dimension $D \times D$ while $x$ is some arbitrary $D$ dimensional vector~\footnote{Frequently one uses a trace instead of projection on $x$. For large $D$ two formulations are the same.}, in our case its components are $x_i=\delta_{i,1}$. When solving the Lindblad equation (\ref{eq:Lin}), $\ket{\rho}$ represents a density operator which is viewed as a member of a Hilbert space of operators, with local basis being Pauli matrices $\sigma^0=\sigma^{\rm x}, \sigma^1=\sigma^{\rm y}, \sigma^2=\sigma^{\rm z}, \sigma^3=\mathbbm{1}$. For a given $\ket{\rho}$ the choice of matrices $M_j^{\nu_j}$ is not unique, however, as we will see, there is a preferred choice corresponding to Schmidt decomposition of $\ket{\rho}$.

\subsection{Evolution}
The above setting is used to solve master equation $\dot{\rho}={\cal L}\rho$, where ${\cal L}$ is some linear operator. In our simulation ${\cal L}$ is just the operator corresponding to the right side of the Lindblad equation (\ref{eq:Lin}). Solution can be formally written as $\rho(t)=\prod{\exp{({\cal L}\Delta t)}}\, \rho(0)$ in terms of propagators $\exp{({\cal L}\Delta t)}$ for a small time step $\Delta t$; in our simulations we use $\Delta t=0.05$. Each small-step propagator is then decomposed according to the Trotter-Suzuki formula. Writing ${\cal L}={\cal A}+{\cal B}$ as a sum of two parts, each of which is a sum of mutually commuting terms, we have
\begin{equation}
\fl 
\exp{({\cal L}\Delta t+{\cal O}(\Delta t^{k+1}))}\approx \exp{(a_1 {\cal A} \Delta t)} \exp{(b_1 {\cal B} \Delta t)} \exp{(a_2 {\cal A} \Delta t)} \cdots \exp{(b_{k} {\cal B} \Delta t)}.
\label{eq:trotter}
\end{equation}  
We use a fourth order ($k=4$) decomposition~\cite{FR} with $a_1=s/2,\, b_1=s,\, a_2=(1-s)/2,\, b_2=(1-2s),\, a_3=(1-s)/2,\, b_3=s,\, a_4=s/2,\, b_4=0$, where $s=1/(2-\sqrt[3]{2})$. In our Lindblad equation we have only nearest neighbor unitary terms and at most two-qubit nearest neighbor dissipative terms. We put into ${\cal A}$ all even bonds, into ${\cal B}$ all odd bonds,  while single qubits terms are equally split between the two terms. ${\cal A}$ and ${\cal B}$ are therefore sums of commuting two-qubit terms. Basic operation we have to do is then $\exp{({\cal A}_{j,j+1} \epsilon )}$, where $\epsilon \propto \Delta t$. For time-independent ${\cal A}_{j,j+1}$ this is a fixed linear operator on the Hilbert space of operators. Its matrix representation $A_{j,j+1}$ in the basis $\ket{\sigma_j^{\nu_j}}$ can be calculated in advance, 
\begin{equation}
[A_{j,j+1}]_{\tilde{\nu}_j \tilde{\nu}_{j+1},\nu_j \nu_{j+1}}=\tr{( \sigma_j^{\tilde{\nu}_j} \sigma_{j+1}^{\tilde{\nu}_{j+1}} \exp{({\cal A}_{j,j+1} \epsilon )} \sigma_j^{\nu_j} \sigma_{j+1}^{\nu_{j+1}})}/4.
\end{equation}
$A_{j,j+1}$ is orthogonal for unitary $\exp{({\cal A}_{j,j+1} \epsilon )}$.

Every bipartite state can be written in the form of the Schmidt decomposition. Splitting a system into the first $j$ sites ($1,\ldots,j$) and the rest, we can write
\begin{equation}
\ket{\rho}=\sum_{\alpha} \lambda^{(j)}_\alpha \ket{w^{{\rm L} (j)}_\alpha} \otimes \ket{w^{{\rm R} (j)}_\alpha},
\end{equation}
with an explicit form of orthogonal states $\ket{w^{{\rm L}(j)}_\alpha}$ and $\ket{w^{{\rm R}(j)}_\alpha}$,
\begin{eqnarray}
\ket{w^{{\rm L} (j)}_\alpha} &=& \sum_{\beta,\nu_1,\ldots,\nu_j}{ x^*_\beta [M_1^{\nu_1}\cdots M_j^{\nu_j}]_{\beta,\alpha} \frac{1}{\lambda^{(j)}_\alpha} \ket{\sigma_1^{\nu_1} \cdots \sigma_j^{\nu_j}}}  \nonumber \\
\ket{w^{{\rm R} (j)}_\alpha} &=& \sum_{\beta,\nu_{j+1},\ldots,\nu_n}{[M_{j+1}^{\nu_{j+1}}\cdots M_n^{\nu_n}]_{\alpha,\beta}\, x_\beta \ket{\sigma_{j+1}^{\nu_{j+1}} \cdots \sigma_{n}^{\nu_n}}} .
\end{eqnarray}
In terms of matrices $M^{\nu_j}_j$ the orthogonality of $\ket{w^{{\rm L}(j)}_\alpha}$ is reflected in the condition
\begin{equation}
\sum_{\nu_j}{ \{M^{\nu_j}_j\}^\dagger {\rm diag}(\{ \lambda^{(j-1)}\}^2) M^{\nu_j}_j }={\rm diag}( \{ \lambda^{(j)}\}^2),
\label{eq:A1}
\end{equation}
for all $j$, while for $\ket{w^{{\rm R}(j)}_\alpha}$
\begin{equation}
\sum_{\nu_j}{M^{\nu_j}_j \{M^{\nu_j}_j\}^\dagger}=\mathbbm{1},
\label{eq:A2}
\end{equation}
must hold. Coefficients $\lambda^{(j)}_\alpha$ are called the Schmidt coefficients. How well the tDMRG method works crucially depends on how fast the ordered $\lambda^{(j)}_\alpha$ decrease with $\alpha$. In general time evolution can not be done in an exact way unless we increase the dimension of matrices $M_j^{\nu_j}$ exponentially with time. To keep dimensions fixed we have to make a truncation. This is done in an optimal way (in terms of probabilities $(\lambda^{(j)}_\alpha)^2$) if we make sure that we drop states corresponding to the smallest Schmidt coefficients. To do this we first calculate the action of $A_{j,j+1}$ on the matrix product operator (MPO) form (\ref{eq:MPO}); it transforms a product of two matrices $M^{\nu_j}_j$ and $M^{\nu_{j+1}}_{j+1}$ at sites $j$ and $j+1$ into a single bigger matrix $\Theta$,
\begin{equation}
\Theta_{\nu_j \alpha,\nu_{j+1} \beta}=\lambda_{\alpha}^{(j-1)} \sum_{\tilde{\nu}_j,\tilde{\nu}_{j+1}}{[A_{j,j+1}]_{\nu_j \nu_{j+1},\tilde{\nu}_j \tilde{\nu}_{j+1}} \, [ M^{\tilde{\nu}_j}_j M^{\tilde{\nu}_{j+1}}_{j+1}]_{\alpha,\beta}}.
\end{equation}
The reason to multiply it with $\lambda^{(j-1)}_\alpha$ is to recover Schmidt decomposition after doing a singular value decomposition on $\Theta$, decomposing it as $\Theta=U d V$, with unitary $U$ and $V$ and diagonal $d$. It turns out that if we prescribe new matrices $\tilde{M}$ and new coefficients $\tilde{\lambda}^{(j)}_\gamma$ as
\begin{eqnarray}
\tilde{\lambda}^{(j)}_\gamma &=& d_\gamma, \nonumber \\
 \left[ \tilde{M}^{\nu_j}_j \right]_{\alpha,\gamma} &=&  U_{\nu_j \alpha,\gamma} 
\frac{\tilde{\lambda}_{\gamma}^{(j)}}{\lambda^{(j-1)}_\alpha}, \nonumber \\
\left[ \tilde{M}^{\nu_{j+1}}_{j+1} \right]_{\gamma,\alpha} &=&  V_{\gamma,\nu_{j+1}\alpha},
\end{eqnarray} 
and if $\exp{({\cal A}_{j,j+1} \epsilon )}$ is unitary, orthogonality conditions (\ref{eq:A1},\ref{eq:A2}) are preserved. The whole Trotter-Suzuki step (\ref{eq:trotter}), advancing in time by $\Delta t$, takes ${\cal O}(kn(dD)^3)$ operations.

\subsection{Reorthogonalization of MPO}
One Trotter-Suzuki step is optimal if all transformations preserve Schmidt decomposition, i.e., if they are unitary. Coupling to the bath and dephasing obviously violate this condition. Because the bath part ${\cal L}^{\rm bath}$ destroys unitarity only at the two boundaries we included it in ${\cal A}$ and ${\cal B}$ terms. Dephasing part though, acting on each spin, would affect unitarity also in the bulk of the chain. As a matter of numerical convenience we have not included it in the Trotter-Suzuki decomposed part (\ref{eq:trotter}) but have instead applied it only at the end of each Trotter-Suzuki step as $\exp{({\cal L}^{\rm deph}\Delta t)}$. This corresponds to dephasing acting in a kicked manner and not in a continuous way and has no physical consequences for our results. 

After each step of length $\Delta t$, we are left with a MPO $\ket{\rho}$ whose matrices do not correspond to Schmidt decomposition anymore (due to non-unitary propagators in between). Before proceeding to the next step we reorthogonalize matrices $M_j^{\nu_j}$ in order to recover the optimality of the Schmidt decomposition. This takes of order ${\cal O}(ndD^3)$ steps and is done in the following way~\cite{Prosen,Shi}. First, we recursively construct matrices $V^{\rm L}_j$ and $V^{\rm R}_j$ as
\begin{eqnarray}
V^{\rm L}_{j}&=&\sum_{\nu_j}{(M^{\nu_j}_j)^\dagger V^{\rm L}_{j-1} M^{\nu_j}_j },\quad j=1,\ldots,n-1\\
V^{\rm R}_{j-1}&=&\sum_{\nu_j}{M^{\nu_j}_j V^{\rm R}_{j} (M^{\nu_j}_j)^\dagger },\quad j=n,\ldots,2,
\end{eqnarray}
starting with the matrix $[V^{\rm L}_0]_{k,l}=x_k x^*_l$ and $[V^{\rm R}_n]_{k,l}=x_k x^*_l$ at the boundaries. If conditions (\ref{eq:A1},\ref{eq:A2}) are satisfied the matrices $V^{\rm L}_j V^{\rm R}_j$ are diagonal. In general this is not the case so we have to rotate each $M_j^{\nu_j}$ in order to recover orthogonality. To do this we calculate a square root of a non-negative $V^{\rm L}_j$ and diagonalize the matrix $W=\sqrt{V^{\rm L}_j}V^{\rm R}_j \sqrt{V^{\rm L}_j}=U\,d\,U^\dagger$, in terms of diagonal $d$ and unitary $U$. Finally, we calculate unitary $G_i=d^{-1/2} U^\dagger (V^{\rm L}_j)^{1/2}$ and its inverse $G_i^{-1}=(V^{\rm L}_j)^{-1/2} U d^{1/2}$ for $j=1,\ldots,n-1$. We also set $G_0=G_n=\mathbbm{1}$. New matrices $\tilde{M}_j^{\nu_j}$ in MPO (\ref{eq:MPO}), respecting Schmidt decomposition, are then obtained by rotations
\begin{equation}
\tilde{M}_j^{\nu_j}=G_{j-1} M_j^{\nu_j} G_j^{-1},\quad j=1,\ldots,n.
\end{equation}

\end{document}